\documentclass[12pt]{article}
\usepackage{axodraw}
\usepackage{cite}
\begin{document}
\newcommand{\beq}{\begin{equation}}
\newcommand{\eeq}{\end{equation}}
\newcommand{\beqa}{\begin{eqnarray}}
\newcommand{\eeqa}{\end{eqnarray}}
\newcommand{\beqar}{\begin{eqnarray*}}
\newcommand{\eeqar}{\end{eqnarray*}}
\newcommand{\ben}{\begin{enumerate}}
\newcommand{\een}{\end{enumerate}}
\newcommand{\bit}{\begin{itemize}}
\newcommand{\eit}{\end{itemize}}
\newcommand{\al}{\alpha}
\newcommand{\be}{\beta}
\newcommand{\del}{\delta}
\newcommand{\D}{\Delta}
\newcommand{\eps}{\epsilon}
\newcommand{\ga}{\gamma}
\newcommand{\Ga}{\Gamma}
\newcommand{\ka}{\kappa}
\newcommand{\inn}{\!\cdot\!}
\newcommand{\h}{\eta}
\newcommand{\kk}{\varphi}
\newcommand\F{{}_3F_2}
\newcommand{\la}{\lambda}
\newcommand{\La}{\Lambda}
\newcommand{\na}{\nabla}
\newcommand{\Om}{\Omega}
\newcommand{\p}{\phi}
\newcommand{\sig}{\sigma}
\renewcommand{\t}{\theta}
\newcommand{\z}{\zeta}
\newcommand{\ssc}{\scriptscriptstyle}
\newcommand{\eg}{{\it e.g.,}\ }
\newcommand{\ie}{{\it i.e.,}\ }
\newcommand{\labell}[1]{\label{#1}} 
\newcommand{\reef}[1]{(\ref{#1})}
\newcommand\prt{\partial}
\newcommand\veps{\varepsilon}
\newcommand\ls{\ell_s}
\newcommand\cF{{\cal F}}
\newcommand\cA{{\cal A}}
\newcommand\cS{{\cal S}}
\newcommand\cT{{\cal T}}
\newcommand\cC{{\cal C}}
\newcommand\cL{{\cal L}}
\newcommand\cM{{\cal M}}
\newcommand\cN{{\cal N}}
\newcommand\cG{{\cal G}}
\newcommand\cI{{\cal I}}
\newcommand\cJ{{\cal J}}
\newcommand\cl{{\iota}}
\newcommand\cP{{\cal P}}
\newcommand\cQ{{\cal Q}}
\newcommand\cg{{\it g}}
\newcommand\cR{{\cal R}}
\newcommand\cB{{\cal B}}
\newcommand\cO{{\cal O}}
\newcommand\cK{{\cal K}}
\newcommand\cH{{\cal H}}
\newcommand\tcO{{\tilde {{\cal O}}}}
\newcommand\bz{\bar{z}}
\newcommand\bw{\bar{w}}
\newcommand\hF{\hat{F}}
\newcommand\hA{\hat{A}}
\newcommand\hT{\hat{T}}
\newcommand\htau{\hat{\tau}}
\newcommand\hD{\hat{D}}
\newcommand\hf{\hat{f}}
\newcommand\hg{\hat{g}}
\newcommand\hp{\hat{\phi}}
\newcommand\hi{\hat{i}}
\newcommand\ha{\hat{a}}
\newcommand\hQ{\hat{Q}}
\newcommand\hP{\hat{\Phi}}
\newcommand\hS{\hat{S}}
\newcommand\hX{\hat{X}}
\newcommand\tL{\tilde{\cal L}}
\newcommand\hL{\hat{\cal L}}
\newcommand\tG{{\widetilde G}}
\newcommand\tg{{\widetilde g}}
\newcommand\tphi{{\widetilde \phi}}
\newcommand\tPhi{{\widetilde \Phi}}
\newcommand\te{{\tilde e}}
\newcommand\tk{{\tilde k}}
\newcommand\tf{{\tilde f}}
\newcommand\tF{{\widetilde F}}
\newcommand\tK{{\widetilde K}}
\newcommand\tE{{\widetilde E}}
\newcommand\tpsi{{\tilde \psi}}
\newcommand\tX{{\widetilde X}}
\newcommand\tD{{\widetilde D}}
\newcommand\tO{{\widetilde O}}
\newcommand\tS{{\tilde S}}
\newcommand\tB{{\widetilde B}}
\newcommand\tA{{\widetilde A}}
\newcommand\tT{{\widetilde T}}
\newcommand\tC{{\widetilde C}}
\newcommand\tV{{\widetilde V}}
\newcommand\thF{{\widetilde {\hat {F}}}}
\newcommand\Tr{{\rm Tr}}
\newcommand\tr{{\rm tr}}
\newcommand\STr{{\rm STr}}
\newcommand\M[2]{M^{#1}{}_{#2}}
\newcommand\nn{\nonumber}
\newcommand\lan{\langle}
\newcommand\ran{\rangle}
\parskip 0.3cm

\vspace*{1cm}

\begin{center}
{\bf \Large Higher-derivative corrections   to type II supergravity:\\Four Ramond-Ramond terms}

\vspace*{1cm}

{Hamid R. Bakhtiarizadeh\footnote{hamidreza.bakhtiarizadeh@stu-mail.um.ac.ir} and {Mohammad R. Garousi\footnote{garousi@um.ac.ir}}}\\
\vspace*{1cm}
{ \it Department of Physics, Ferdowsi University of Mashhad,\\ P.O. Box 1436, Mashhad, Iran }

\vskip 0.6 cm

\vspace{2cm}
\end{center}

\begin{abstract}
\baselineskip=18pt
It is known that the sphere-level  S-matrix element of four type II superstrings has one    kinematic factor. At the low energy limit, this factor produces the kinematic factor of  the corresponding Feynman amplitudes in the supergravity.  It also produces   higher-derivative couplings of four strings. In this paper,     we explicitly calculate  the kinematic factor of four   RR states in the supergravity. Using this factor,    we then find the  eight-derivative P-even and P-odd couplings of four RR fields, including the self-dual RR five-form field strength. We show that  the P-even couplings are mapped  to  the standard $\bar{R}^4$ couplings  by linear T-duality and S-duality transformations. We also confirm the P-even couplings with  direct calculations   in  type II superstring theories. 
\end{abstract}
\setcounter{page}{0}
\setcounter{footnote}{0}
\newpage

\section{Introduction}

Superstring theories at low energy limit are described appropriately by  supergravities  which include only the massless modes and their interactions at two-derivative level. These theories inherit many symmetries of the superstring theories such as string dualities \cite{Alvarez:1994dn,Giveon:1994fu,Sen:1998kr,Vafa:1997pm }. For many purposes, it is enough to use only these effective theories, but there are   situations for which one must go beyond the lowest order terms in the effective actions.   The   higher order terms must be  corrections in  $ \alpha' $ and in the string coupling constant $ g_{s} $. The main challenge thus is to implement the symmetries of the superstring theories to find an effective action that incorporates  all such corrections, including  non-perturbative effects \cite{Green:1997tv}.

Subleading terms in type II effective actions start at  eight derivative level, and were first calculated at the tree level from four-graviton scattering \cite{Schwarz:1982jn,Gross:1986iv} as well as from the $ \sigma $-model beta function  \cite{Grisaru:1986px,Grisaru:1986dk,Grisaru:1986kw,Freeman:1986br,Grisaru:1986vi,Freeman:1986zh}. They take the following   form at tree level in string frame:
\beqa
S\supset \frac{\gamma }{3.2^7\kappa^2} \int d^{10}x \, e^{-2\phi} \sqrt{-G}\left(t_8t_8R^4+\frac{1}{4}\eps_{8}\eps_{8}R^4\right),\labell{t8t8}
\eeqa
where $\gamma=\frac{\alpha'^3\z(3)}{2^{5}}$ and $t_8$ is a tensor which is antisymmetric within a pair of indices and is symmetric under exchange of the pair of indices. The above expression, however, cannot be complete, as supersymmetry will necessarily bring in additional higher order terms built from the other fields in the supergravity multiplet. This includes the B-field and dilaton in the NSNS sector, the n-form field strengths in the RR sector, and their corresponding fermionic superpartners. It would be desirable to obtain a   supersymmetric invariant action at the eight-derivative level which is   the completion of the above terms \cite{deRoo:1992zp,Howe:1983sra,Nilsson:1986rh,Green:1998by,Peeters:2000qj,Peeters:2001ub,Liu:2013dna}.

The bosonic couplings in the effective action \reef{t8t8} may also be found by constraining  it  to be consistent with the string dualities  \cite{Garousi:2012yr,Garousi:2012jp,Garousi:2013lja,Godazgar:2013bja,Garousi:2013qka,Liu:2013dna}.  The couplings at weak field level, \ie four-field couplings,  may  also be found  more directly from   the corresponding scattering amplitude of four vertex operators. They must be also consistent with linear string dualities. The sphere-level scattering amplitude of  four strings has the following structure in  Einstein frame \cite{Schwarz:1982jn,Gross:1986iv}:
\beqa
\cA&=&  \left(\frac{\Gamma(-e^{-\phi_0/2}s/8)\Gamma(-e^{-\phi_0/2}t/8)\Gamma(-e^{-\phi_0/2}u/8)}{\Gamma(1+e^{-\phi_0/2}s/8)\Gamma(1+e^{-\phi_0/2}t/8)\Gamma(1+e^{-\phi_0/2}u/8)} \right)\cK\labell{contact}
\eeqa
where $\cK$ is kinematic factor that depends on external states, $\phi_0$ is the constant dilaton background and $s,\, t,\, u$ are the Mandelstam variables\footnote{Relation between the Einstein frame metric and the string frame metric   is $G^E_{\mu\nu}=e^{-\phi/2}G^S_{\mu\nu}$.} The low energy expansion of the Gamma functions is 
\beqa
\frac{\Gamma(-e^{-\phi_0/2}s/8)\Gamma(-e^{-\phi_0/2}t/8)\Gamma(-e^{-\phi_0/2}u/8)}{\Gamma(1+e^{-\phi_0/2}s/8)\Gamma(1+e^{-\phi_0/2}t/8)\Gamma(1+e^{-\phi_0/2}u/8)}&=&-\frac{2^9e^{3\phi_0/2}}{stu}-2\zeta(3) +\cdots\labell{expa}
\eeqa
where dots refer to higher order contact terms. Thus, the kinematic factor plays two roles. It produces the Feynman amplitude of four massless strings in the supergravity \cite{Sannan:1986tz}.  On the other hand, it produces the couplings of four strings at order $\alpha'^3$ \cite{Schwarz:1982jn,Gross:1986iv}.

The kinematic factor of RR states   involves various traces over the 10-dimensional gamma matrices. Performing the traces, one expects that the amplitudes at two-momentum level are reproduced by the corresponding Feynman amplitudes in the supergravity, and   at eight-momentum level  they reproduce  the eight-derivative couplings in the action \reef{t8t8}. Such calculation for the scattering amplitude of two RR and two NSNS states has been done explicitly in \cite{Bakhtiarizadeh:2013zia}. It has been shown that the couplings at eight-derivative level are related to four NSNS couplings found in \cite{Schwarz:1982jn,Gross:1986iv} through the linear T-duality and S-duality \cite{Bakhtiarizadeh:2013zia}.

In this paper, we are interested in the couplings of four RR fields in the effective action \reef{t8t8}, including the RR five-form field strength which must be  self-dual. We use the above double roles of the kinematic factor. That is, we first calculate the kinematic factor of four RR states in the type II supergravities, then we use it to find the couplings of four RR states at order $\alpha'^3$. The standard type IIB supergravity, however,  is off by the fact that it does not include the self-duality of the  RR five-form field strength. The self-duality must be imposed  by hand  on equations of motion \cite{Bergshoeff:1995sq}. In this paper, we impose   the self-duality of the RR five-form field strength by hand in  the scattering amplitudes. The couplings   we have found then have P-even and P-odd parts. We will  confirm the P-even couplings  by demonstrating that they are related  to four NSNS couplings \cite{Schwarz:1982jn,Gross:1986iv,Gross:1986mw} through the linear T-duality and S-duality transformations. We will also confirm them by direct comparison with the kinematic factor in the type II superstring theories.

The paper is arranged as follows: In section \ref{Field theory amplitude}, we use  the type II   supergravities to calculate  various  scattering amplitudes of four   RR states, and find their corresponding kinematic factors.    We then transform these  factors to spacetime and find various   couplings of four RR field strengths at order $\alpha'^3$. After imposing the self-duality on the RR five-form field strength, we find all  P-even and P-odd couplings. In section 3, the S-duality and  T-duality have been used as guiding principles to find the P-even couplings of four RR field strengths from the sphere-level couplings of four NSNS states. We find exact agreement with the P-even part of the above couplings. In section \ref{String theory amplitude}, we confirm the P-even    couplings directly in type II superstring theories by performing the traces in the corresponding  kinematic factor of the S-matrix element of four RR vertex operators in the RNS   and in  the  Pure spinor formalisms. 

\section{Field theory amplitude }\label{Field theory amplitude}

In this section we are going to calculate the S-matrix elements of four RR fields in   supergravity. These amplitudes have the following structure in the Einstein  frame:
\beqa
A&=&   \frac{\cK_s}{s}+\frac{\cK_t}{t}+\frac{\cK_u}{u} \nn\\
&=& \frac{1}{stu}\bigg(tu\cK_s+su\cK_t+st\cK_u\bigg)
\eeqa
 where $\cK_s,\, \cK_t$ and $\cK_u$ are the field theory kinematic factors in $s$-, $ t$-  and $u$-channel, respectively.  The Mandelstam variables are defined as $ s = -4\alpha' k_1\inn k_2 $, $ u = -4\alpha' k_1\inn k_3 $, $ t = -4\alpha' k_2\inn k_3 $ and they satisfy the on-shell condition $ s + t + u = 0 $. Comparing this amplitude with the leading term of the string theory amplitude \reef{contact}, one finds the following relation between the field theory and the string theory kinematic factors:
\beqa
 \cK=-2^{-9} e^{-3\phi_0/2}\bigg(tu\cK_s+su\cK_t+st\cK_u\bigg)\labell{kin}
\eeqa
 Multiplying this factor by $-2\zeta(3) $ and transforming it   to the the coordinate space, one then  finds the couplings of four RR fields at order $\alpha'^3$.

The type II supergravities  describe interactions of   massless fields of type II superstring theories at two-derivative level.  The type IIA supergravity in the Einstein frame is given as (see \eg, \cite{Becker:2007})
\beqa
S_{IIA}&=&\frac{1}{2\kappa^2} \int d^{10}x \sqrt{-G} \left(R-\frac{1}{2} \prt_{\mu} \Phi \prt^{\mu} \Phi-\frac{1}{2} e^{-\Phi} |H|^2-\frac{1}{2} \sum_{n=2,4}e^{\frac{5-n}{2}\Phi} |\tilde{F}^{(n)}|^2\right)\nonumber\\
&&-\frac{1}{4\kappa^2} \int\, B\wedge dC^{(3)}\wedge dC^{(3)},\labell{IIA}
\eeqa
where $ R $ is the scalar curvature, $ \Phi $ is the dilaton field and $H$ is the B-field strength  $H=dB$. The RR field strengths are $\tilde{F}^{(2)}=dC^{(1)}$ and $\tilde{F}^{(4)}=dC^{(3)}-   H\wedge C^{(1)}$. The above action is the   reduction of  11-dimensional supergravity on manifold $R^{1,9}\times S^1$.

 Unlike the type IIA supergravity, there is a challenging feature in type IIB supergravity which is  the self-duality of  five-form field strength. It is hard to  formulate the action in a manifestly covariant form.    One way to find the action is to first construct the supersymmetric equations of motion, and then to write down an action that reproduces those equations when the self-duality condition is imposed by hand. 
The type IIB supergravity in the Einstein frame is given as (see \eg, \cite{Becker:2007})
\beqa
S_{IIB}&=&\frac{1}{2\kappa^2} \int d^{10}x \sqrt{-G} \left(R-\frac{1}{2} \prt_{\mu} \Phi \prt^{\mu} \Phi-\frac{1}{2} e^{-\Phi} |H|^2-\frac{1}{2\alpha} \sum_{n=1,3,5} e^{\frac{5-n}{2}\Phi} |\tilde{F}^{(n)}|^2\right)\nonumber\\
&&-\frac{1}{4\kappa^2} \int\,  H\wedge dC^{(2)}\wedge C^{(4)},\labell{IIB}
\eeqa
where $\alpha=1$ for $n=1,3$ and  $\alpha=2$ for $n=5$. The RR field strengths in this case are $\tilde{F}^{(1)}=dC^{(0)}$,   $\tilde{F}^{(3)}=dC^{(2)}- H C^{(0)}$ and 
\beqa
\tilde{F}^{(5)}=dC^{(4)}-\frac{1}{2} C^{(2)} \wedge H+ \frac{1}{2} B \wedge dC^{(2)} . \labell{tildeF5}
\eeqa
The self-duality condition that must be imposed  in the equations of motion by hand, is
 \beqa
\tilde{F}^{(5)}=\star \tilde{F}^{(5)} . \labell{selfdual}
\eeqa
   We will show that without the above self-duality condition, the action \reef{IIB} does not reproduce correctly the S-matrix element of string theory at low energy. However, imposing this constraint by hand on the S-matrix elements, we will find the consistency between field theory and string theory S-matrix elements.

Using  the above supergravity actions, one can read various vertices, propagators, and accordingly calculate the Feynman amplitude of four RR states. For this purpose, we assume  the massless fields are small perturbations around the flat background, \ie 
\beqa
g_{\mu\nu}=\eta_{\mu\nu}+2\kappa h_{\mu\nu};\,B^{(2)}=2\kappa b^{(2)}\,;\,\,
\Phi\,=\,\phi_0+\sqrt{2}\kappa \phi\; \labell{perturb} 
\eeqa  
The explicit form of the propagators and the vertices that we need in this paper appears in Appendix A. 
  The external states satisfy the on-shell relations $k^2=0$ and $k\inn \veps=0$ where $\veps^{\mu_1\mu_2\cdots}$ is the polarization of external RR states. Therefore, the couplings that we will find does not contain $\prt_{\mu}F^{\mu\mu_1\mu_2\cdots}$.

\subsection{$\prt F^{(n)}\prt F^{(n)}\prt F^{(n)}\prt F^{(n)}$ couplings}

There are five types of couplings in this section, \ie $n=1,2,3,4,5$. When the   four RR  forms have the same rank, the actions \reef{IIA} and \reef{IIB} dictate that for the cases  $ n=1,2,3 $, the Feynman amplitude in the $s$-channel    is given by the following expression: 
\beqa 
A_s&=&\left[\tV_{F_1^{(n)}F_2^{(n)}h}\right]^{\mu \nu}\left[\tG_{h} \right]_{\mu \nu, \lambda\rho}\left[\tV_{hF_3^{(n)}F_4^{(n)}}\right]^{\lambda\rho}+\tV_{F_1^{(n)}F_2^{(n)}\phi} \tG_{\phi}\tV_{\phi F_3^{(n)}F_4^{(n)}} , \labell{s-chaFnFnFnFn}
\eeqa
where   the vertices and propagators are given in the Appendix A.  
The amplitude  in the $u$-channel is the same as $A_s$ in which the particle labels of the   RR fields are interchanged, \ie $A_u= A_s(2 \leftrightarrow 3)$. Similarly, the amplitude in the $t$-channel is the same as $A_u$ in which the particle labels of the external RR fields are interchanged, \ie $A_t= A_u(3 \leftrightarrow 4)$.

Replacing the   vertices and propagators in \reef{s-chaFnFnFnFn}, one can calculate the string theory kinematic factor \reef{kin}.
To convert this factor to the couplings in the form of $(\prt F)^4$, we use the conservation of momentum, $ \sum_{i=1}^{4}k_i=0 $, and the on-shell relations on the external states to write the multiples of two Mandelstam variables which appear in \reef{kin}, as  
\beqa
st&=&8\alpha'^2\left(k_1\inn k_3 \, k_2\inn k_4-k_1\inn k_2 \, k_3\inn k_4-k_1\inn k_4 \, k_2\inn k_3\right),\nn\\
su&=&8\alpha'^2\left(k_1\inn k_4 \, k_2\inn k_3-k_1\inn k_2 \, k_3\inn k_4-k_1\inn k_3 \, k_2\inn k_4\right),\nn\\
tu&=&8\alpha'^2\left(k_1\inn k_2 \, k_3\inn k_4-k_1\inn k_3 \, k_2\inn k_4-k_1\inn k_4 \, k_2\inn k_3\right),\labell{identities}
\eeqa
where on the right-hand side each label appears once in each term.   
With the assistance of a field-theory inspired package for Mathematica, ``xTras'' \cite{Nutma:2013zea} as well as a symbolic computer algebra system for field theory problems known as ``Cadabra" \cite{Peeters:2006kp,Peeters:2007wn}, we find the following couplings for $n=1,2,3$ in the Einstein frame\footnote{All contracted indices have been written as subscripts for easier readability.}:
\beqa
\cK&\!\!\!\!=\!\!\!\!& -\frac{\alpha'^3e^{5\phi_0/2}}{  2^{9}\kappa^2}  \left[ 6 F_{a,c} F_{b,d}F_{a,b} F_{c,d} -F_{a,c} F_{b,d}F_{a,c} F_{b,d} \right]\labell{K123}\\
\cK&\!\!\!\!=\!\!\!\!&
  \frac{\alpha'^3e^{3\phi_0/2}}{  2^{11}\kappa^2}   \left[ 8 F_{ab,e} F_{bc,f} F_{ad,f} F_{cd,e} -2 F_{ab,e}F_{ab,f} F_{cd,f} F_{cd,e}+F_{ab,e}F_{ab,e} F_{cd,f} F_{cd,f} \right] \nonumber\\
\cK&\!\!\!\!=\!\!\!\!& -\frac{\alpha'^3e^{\phi_0/2}}{  2^{11}3^2\kappa^2}    \big[ 18 F_{abc,g}F_{bcd,h} F_{aef,h} F_{def,g} -2 F_{abc,g}F_{abc,h} F_{def,h} F_{def,g}+18 F_{abc,g}F_{bcd,h} F_{aef,g} F_{def,h}\nn\\&&\qquad\qquad\qquad\qquad\qquad\qquad-18 F_{abc,g}F_{bcd,g} F_{aef,h} F_{def,h}+F_{abc,g}F_{abc,g} F_{def,h} F_{def,h} \big] \nn .
\eeqa
The antisymmetric properties of the RR field strengths have been taken into account to simplify the kinematic factors in above form. However, multi-term symmetry, \ie the Bianchi identity for the RR field strength, $dF=0$,  which relates a sum of terms with different index distribution,  has not yet been taken into account. This identity reduces the number of couplings to the minimal number.

To do this last step, we use the following algorithm: The general structure of each coupling in the momentum space contains four RR field strengths that each one caries one momentum index. We first write it in terms of independent variables. This can be done by writing the RR field strengths in terms of RR potentials and using the conservation of momentum and the on-shell relations to rewrite the coupling in terms of independent variables, \ie writing $k_4=-k_3-k_2-k_1$ and $k_3\inn \veps_4=-k_1\inn \veps_4-k_2\inn\veps_4$.  This impose all symmetries, including the Bianchi identity. Then, we consider all possible   contractions of four RR field strengths with unknown coefficients and rewrite them  in terms of independent variables. By comparing these two results, one finds some algebraic equations  between the unknown coefficients which can be solved to find the coefficients.

To find the minimum number of couplings, we set all unknown coefficients to zero except one of them and solve the equations.    If there is a solution, then the   coefficient of the minimum terms which is one in this case,   would be found.   Otherwise, we have to repeat this procedure by setting all  coefficients to zero   except two of them.  If there is a solution, then the   coefficient of the minimum terms which is two in this case,   would be found.  We continue this approach to find  the minimal number of couplings.

Performing this calculation for the couplings \reef{K123}, we simplify them to the following couplings in the  string frame: 
\beqa
S &\supset &\frac{\gamma}{8\kappa^2}\int d^{10}x\, e^{2\phi_0} \sqrt{-G}  \bigg[2F_{a,b} F_{a,b} F_{c,d} F_{c,d} \labell{F3F3F3F3}\\
&&+ 2F_{a d ,e} F_{a b ,c} F_{c f ,d} F_{e f ,b} + 2F_{a d ,e} F_{a b ,c} F_{c f ,b} F_{e f ,d}- 
\frac{1}{2} F_{a b ,d} F_{a b ,c} F_{e f ,d} F_{e f ,c}   \nonumber\\
&&+ F_{a e f ,d} F_{a b c ,d} F_{b e g ,h} F_{c f h ,g}  + F_{a b e ,f} F_{a b c ,d} F_{c f g ,h} F_{d e g ,h} + F_{a b e ,f} F_{a b c ,d} F_{c g h ,e} F_{f g h ,d} \bigg] \nn.
\eeqa
While the above algorithm reduces the number of terms  for $n=1,3$, it does not reduce the three couplings in the case of $n=2$. However, the index distribution  is changed. It means there are at least two different index distributions for the three terms that are identical up to the Bianchi identity. Note that to find the standard sphere-level dilaton factor $e^{-2\phi_0}$ in the string frame, one has to normalize the RR potential $C$ with $e^{\phi_0}C$. The normalization of the RR fields in above action is consistent with the supergravities \reef{IIA} and \reef{IIB}.

For $n=4$ case, there is another contribution to the scattering amplitude in the $s$-channel  which is coming from the Chern-Simons term in \reef{IIA}. The Feynman amplitude  in this case   is given as 
\beqa 
A_s&=&\left[\tV_{F_1^{(4)}F_2^{(4)}h}\right]^{\mu \nu}\left[\tG_{h} \right]_{\mu \nu, \lambda\rho}\left[\tV_{hF_3^{(4)}F_4^{(4)}}\right]^{\lambda\rho}+\tV_{F_1^{(4)}F_2^{(4)}\phi} \tG_{\phi}\tV_{\phi F_3^{(4)}F_4^{(4)}}\nn\\&&+\left[\tV_{\epsilon_{10}F_1^{(4)}F_2^{(4)}b}\right]^{\mu \nu}\left[\tG_{b} \right]_{\mu \nu, \lambda\rho}\left[\tV_{bF_3^{(4)}F_4^{(4)}\epsilon_{10}}\right]^{\lambda\rho} \nn
\eeqa
The amplitude in the first line is the same as the amplitude \reef{s-chaFnFnFnFn}. The term in  the second line has two Levi-Civita tensors which can be replaced by the generalized kronecker delta according to the following expression:
\beqa
\epsilon^{m_1 \cdots m_{d}}\epsilon_{n_1 \cdots n_{d}}= -\delta_{[n_1}{}^{m1} \cdots \delta_{n_d]}{}^{md}\labell{iden},
\eeqa
The massless pole in the $u$-channel is the same as $A_s$ in which the particle labels of the external RR fields are interchanged, \ie $A_u= A_s(2 \leftrightarrow 3)$. Similarly, the massless pole in the $t$-channel is the same as $A_u$ in which the particle labels of the external RR fields are interchanged, \ie $A_t= A_u(3 \leftrightarrow 4)$.


Replacing the   vertices and propagators in above amplitude, one can evaluate the kinematic factor $\cK$. Using the antisymmetry property of the RR field strength,  we simplify the result to the following couplings in the string frame:  
\beqa
S &\supset& -\frac{\gamma}{2^{9}.3^2\kappa^2} \int d^{10}x\,  e^{2\phi_0} \sqrt{-G} \, \bigg[72 F_{a b f g ,e} F_{a b c d ,e} F_{c d r t ,h} F_{f g r t ,h} \labell{F4F4F4F4}\\&&\qquad\qquad\qquad\quad-36 F_{a b f g ,r} F_{a b c d ,e} F_{c d t h ,r} F_{f g t h ,e} -64 F_{a b c f ,g} F_{a b c d ,e} F_{d r t h ,e} F_{f r t h ,g} \nn\\&&\qquad\qquad\qquad\quad-F_{a b c d ,e} F_{a b c d ,e} F_{f g r t ,h} F_{f g r t ,h} +6 F_{a b c d ,f} F_{a b c d ,e} F_{g r t h ,f} F_{g r t h ,e} \bigg].\nn
\eeqa
In this case we have tried to use the Bianchi identity to reduce the number of terms. However, we could not reduce the number to less than five terms. So the above terms are the minimum number of terms for the couplings with structure $(\prt F^{(4)})^4$.

For $n=5$ case, the supergravity action \reef{IIB} dictates that there is only one  contribution to the scattering amplitude in the $s$-channel. The Feynman amplitude in this case is given   as  
\beqa 
A_s&=&\left[\tV_{F_1^{(5)}F_2^{(5)}h}\right]^{\mu \nu}\left[\tG_{h} \right]_{\mu \nu, \lambda\rho}\left[\tV_{hF_3^{(5)}F_4^{(5)}}\right]^{\lambda\rho} \nn
\eeqa
The massless pole in the $u$-channel is the same as $A_s$ in which the particle labels of the external RR fields are interchanged, \ie $A_u= A_s(2 \leftrightarrow 3)$. Similarly, the massless pole in the $t$-channel is the same as $A_u$ in which the particle labels of the external RR fields are interchanged, \ie $A_t= A_u(3 \leftrightarrow 4)$. These amplitudes produce the following   coupling in the string frame:
\beqa
 &\!\!\!\!\!\!\!\!\!\!\!\!\!\!&-\frac{\gamma}{2^{11}.3^2.5\kappa^2} \int d^{10}x e^{2\phi_0}\sqrt{-G} \, \bigg[F_{a b c d e ,k} F_{a b c d e ,l} F_{f g h i j ,k} F_{f g h i j ,l}+F_{a b c d e ,k} F_{a b c d e ,l} F_{f g h i j ,k} F_{f g h i j ,l}\nn\\ &\!\!\!\!\!\!\!\!\!\!\!\!\!\!&\qquad\qquad\qquad\qquad-F_{a b c d e ,k} F_{a b c d e ,k} F_{f g h i j ,l} F_{f g h i j ,l}+10 F_{a b c d e ,f} F_{a b c d g ,f} F_{e h i j k ,l} F_{g h i j k ,l}\labell{5555}\\ &\!\!\!\!\!\!\!\!\!\!\!\!\!\!&\qquad\qquad\qquad\qquad-10 F_{a b c d e ,f} F_{a b c d g ,h} F_{e i j k l ,h} F_{g i j k l ,f}-10 F_{a b c d e ,f} F_{a b c d g ,h} F_{e i j k l ,f} F_{g i j k l ,h}\bigg]\nn
\eeqa
which are only P-even couplings. We have compared them with the corresponding   scattering amplitude of four RR vertex operators in string theory and found disagreement! This indicates that the type IIB supergravity does not correctly describe the couplings of RR five-form field strength.
However, the supergravity \reef{IIB} is expected to  describe   only the  self-dual part of the RR five-form field strength after imposing the self-duality by hand.   Therefore,  we expect the above couplings  to be   physical  only after imposing the following transformation by hand:  
\beqa
F_5 \rightarrow \frac{1}{2}(F_5 + \star F_5 ),\labell{self-duality}
\eeqa
  Then the   couplings \reef{5555} are expected to be consistent with  the $\alpha'^3$ terms of the corresponding string theory scattering amplitude. 

We impose the above self-duality in the couplings \reef{5555} and use the identity \reef{iden} to rewrite the even number of  Levi-Civita tensors  in terms of metric and odd number of Levi-Civita tensors in terms of one Levi-Civita tensor. Using the  ``Cadabra" \cite{Peeters:2006kp,Peeters:2007wn}, we have found the following result: 
\beqa
S &\!\!\!\!\supset\!\!\!\!& -\frac{\gamma}{2^{17}.3^3.5^2\kappa^2}\int d^{10}x\,e^{2\phi_0}\sqrt{-G} \, \bigg[240\bigg(F_{b c d e f ,a} F_{b c d e f ,a} F_{h i j k l ,g} F_{h i j k l ,g}\labell{epsF5F5F5F5}\\&&\qquad\qquad\qquad\quad+180 F_{b c d e f ,a} F_{b g h i j ,a} F_{c d g h l ,k} F_{e f i j l ,k}-360 F_{b c d e f ,a} F_{b c g h i ,a} F_{d e g k l ,j} F_{f h i k l ,j}\nn\\&&\qquad\qquad\qquad\quad+160 F_{b c d e f ,a} F_{b c d g h ,a} F_{e g j k l ,i} F_{f h j k l ,i}-40 F_{b c d e f ,a} F_{g h i j k ,a} F_{b c d g h ,l} F_{e f i j k ,l}\nn\\&&\qquad\qquad\qquad\quad+10 F_{b c d e f ,a} F_{g h i j k ,a} F_{b c d e g ,l} F_{f h i j k ,l}+2 F_{b c d e f ,a} F_{g h i j k ,a} F_{b c d e f ,l} F_{g h i j k ,l}\nn\\&&\qquad\qquad\qquad\quad-30 F_{b c d e f ,a} F_{b g h i j ,a} F_{c d e f l ,k} F_{g h i j l ,k}+40 F_{b c d e f ,a} F_{b c g h i ,a} F_{d e f k l ,j} F_{g h i k l ,j}\nn\\&&\qquad\qquad\qquad\quad+40 F_{b c d e f ,a} F_{b c d g h ,a} F_{e f j k l ,i} F_{g h j k l ,i}-30 F_{b c d e f ,a} F_{b c d e g ,a} F_{f i j k l ,h} F_{g i j k l ,h}\bigg)\nn\\&&\qquad\qquad\quad\qquad- \epsilon_{a b c d e f g h i j} \bigg(360 F_{a b c d e,r} F_{f g k n p,r} F_{h l m n p,q} F_{i j k l m,q}\nn\\&&\qquad\qquad\qquad\quad-180 F_{a b c d e,r} F_{f g k n p,q} F_{h l m n p,r} F_{i j k l m,q}-160 F_{a b c d e,r} F_{f g h k p,r} F_{i l m n p,q} F_{j k l m n,q}\nn\\&&\qquad\qquad\qquad\quad+20 F_{a b c d e,r} F_{f g h i p,r} F_{j k l m n,q} F_{k l m n p,q}-50 F_{a b c d e,r} F_{f g h n p,r} F_{i j k l m,q} F_{k l m n p,q}\nn\\&&\qquad\qquad\qquad\quad+55 F_{a b c d e,q} F_{f g h i p,r} F_{j k l m n,q} F_{k l m n p,r}+50 F_{a b c d e,r} F_{f g h n p,q} F_{i j k l m,q} F_{k l m n p,r}\nn\\&&\qquad\qquad\qquad\quad-50 F_{a b c d e,q} F_{f g h n p,r} F_{i j k l m,q} F_{k l m n p,r}+F_{a b c d e,q} F_{f g h i j,r} F_{k l m n p,q} F_{k l m n p,r}\bigg)\bigg]\nn
\eeqa
The P-odd couplings above are then produced only by the self-duality transformation \reef{self-duality}. The  above couplings must be invariant under the transformation $F^{(5)}\rightarrow \star F^{(5)}$. So they describe the couplings of four self-dual five-forms at order $\alpha'^3$. Note that there is no P-odd couplings in \reef{F4F4F4F4}, so the above   action must produce no couplings with structure $\epsilon_{10}(\prt F^{(4)})^4$ under the T-duality. It is easy to verify it by noticing that it is impossible to have four RR five-form field strengths in the P-odd part that each one carries one Killing index. 

Since the couplings with structure  $(\prt F^{(5)})^4$ have too many indices, it is hard to find all such couplings with unknown coefficients   with the ``xTras'' package \cite{Nutma:2013zea}.  So we could not apply the algorithm given above equation \reef{F3F3F3F3} to reduce the couplings to the minimum number.

\subsection{$\prt F^{(n)}\prt F^{(n)}\prt F^{(n-2)}\prt F^{(n-2)}$ couplings}

Since the maximum rank of the RR field strength is 5, there are three types of couplings in this section, \ie $n=3,4,5$.  The scattering amplitude in the $s$-channel for the cases $n=3,4$  is given by:
\beqa 
A_s&=&\left[\tV_{F_1^{(n)}F_2^{(n)}h}\right]^{\mu \nu}\left[\tG_{h} \right]_{\mu \nu, \lambda\rho}\left[\tV_{hF_3^{(n-2)}F_4^{(n-2)}}\right]^{\lambda\rho}+\tV_{F_1^{(n)}F_2^{(n)}\phi}\tG_{\phi}\tV_{\phi F_3^{(n-2)}F_4^{(n-2)}} \nn
\eeqa
The scattering amplitude in the   $u$-channel is  given as
\beqa 
A_u&=&\left[\tV_{F_1^{(n)}F_3^{(n-2)}b}\right]^{\mu \nu}\left[\tG_{b} \right]_{\mu \nu, \lambda\rho}\left[\tV_{bF_2^{(n)}F_4^{(n-2)}}\right]^{\lambda\rho} \nn
\eeqa
And the Feynman amplitude in the  $t$-channel is the same as $A_u$ in which the particle labels of the external RR fields are interchanged, \ie  $A_t= A_u(3 \leftrightarrow 4)$. 

Replacing the vertices and propagators in the above amplitudes, one finds the following couplings for $n=3,4$ in the string frame: 
\beqa
S &\!\!\!\!\!\supset \!\!\!\!\!& \frac{\gamma}{4\kappa^2} \int d^{10}x\, e^{2\phi_0}\sqrt{-G} \bigg[ 2 F_{a e f ,c} F_{d e f ,b} F_{a ,b} F_{c ,d}+\frac{1}{3}F_{d e f ,c} F_{d e f ,b} F_{a ,c} F_{a ,b}- \frac{1}{6}F_{c d e ,f} F_{c d e ,f} F_{a ,b} F_{a ,b}   \nn\\&&\qquad\qquad
+ F_{a c g h ,f} F_{b d g h ,e} F_{a b ,c} F_{d e ,f}  -  F_{a d f g ,h} F_{b f g h ,e} F_{a b ,c} F_{d e ,c} +  F_{a b g h ,d} F_{c e f g ,h} F_{a b ,c} F_{d e ,f} \nn\\&&\qquad\qquad  +  F_{a f g h ,d} F_{c e g h ,b} F_{a b ,c} F_{d e ,f} + \frac{2}{3} F_{b f g h ,d} F_{c f g h ,e} F_{a d ,e} F_{a b ,c} \bigg]\labell{F4F4F2F2}
\eeqa
where we have also used the algorithm given above equation \reef{F3F3F3F3} to  reduce the couplings to the minimum number.

For the case $n=5$, the type IIB supergravity \reef{IIB} gives  the following Feynman amplitudes in the $s$-channel and $u$-channel: 
\beqa 
A_s&=&\left[\tV_{F_1^{(5)}F_2^{(5)}h}\right]^{\mu \nu}\left[\tG_{h} \right]_{\mu \nu, \lambda\rho}\left[\tV_{hF_3^{(3)}F_4^{(3)}}\right]^{\lambda\rho}\nn\\
A_u&=&\left[\tV_{F_1^{(5)}F_3^{(3)}b}+\tV_{\epsilon_{10}F_1^{(5)}F_3^{(3)}b}\right]^{\mu \nu}\left[\tG_{b} \right]_{\mu \nu, \lambda\rho}\left[\tV_{bF_2^{(5)}F_4^{(3)}}+\tV_{bF_2^{(5)}F_4^{(3)}\epsilon_{10}}\right]^{\lambda\rho}\nn
\eeqa
The amplitude in the  $t$-channel is the same as $A_u$ in which the particle labels of the external RR fields are interchanged. Replacing the vertices and propagators in the above amplitudes, one finds the following couplings in the string frame:
\beqa
 &\!\!\!\! \!\!\!\!&\frac{\gamma}{2^8.5.3^3\kappa^2} \int d^{10}x\, e^{2\phi_0}\sqrt{-G} \, \bigg[120 F_{a b c ,d} F_{e f g ,d} F_{a b c h i ,j} F_{e f g h i ,j}-180 F_{a b c ,d} F_{a e f ,d} F_{b c g h i ,j} F_{e f g h i ,j}\nn\\&&\qquad\qquad\qquad\quad-90 F_{a b c ,d} F_{a b e ,d} F_{c f g h i ,j} F_{e f g h i ,j}+12 F_{a b c ,d} F_{a b c ,d} F_{e f g h i ,j} F_{e f g h i ,j}\nn\\&&\qquad\qquad\qquad\quad+180 F_{a b c ,d} F_{a e f ,g} F_{b c h i j ,g} F_{e f h i j ,d}-180 F_{a b c ,d} F_{a e f ,g} F_{b c h i j ,d} F_{e f h i j ,g}\nn\\&&\qquad\qquad\qquad\quad+90 F_{a b c ,d} F_{a b e ,f} F_{c g h i j ,f} F_{e g h i j ,d}+270 F_{a b c ,d} F_{a b e ,f} F_{c g h i j ,d} F_{e g h i j ,f}\nn\\&&\qquad\qquad\qquad\quad-36 F_{a b c ,d} F_{a b c ,e} F_{f g h i j ,d} F_{f g h i j ,e}+\epsilon_{a b c d e f g k l m} \bigg(F_{h i j,p} F_{k l m,n} F_{a b c d e,n} F_{f g h i j,p}\nn\\&&\qquad\qquad\qquad\quad-F_{h i j,n} F_{k l m,p} F_{a b c d e,n} F_{f g h i j,p}-F_{h i j,p} F_{k l m,p} F_{a b c d e,n} F_{f g h i j,n}\bigg)\bigg]\nn
\eeqa
which has P-even and P-odd parts.

Since the above couplings involves the RR five form field strength, we have to impose the transformation \reef{self-duality} by hand   to produce correct couplings.   We have found the following couplings for the  self-dual RR five-form:
\beqa
S &\!\!\!\!\supset\!\!\!\!&\frac{\gamma}{2^9.5.3^3\kappa^2} \int d^{10}x\, e^{2\phi_0}\sqrt{-G} \, \bigg[240 F_{a b c ,d} F_{e f g ,d} F_{a b c h i ,j} F_{e f g h i ,j}-360 F_{a b c ,d} F_{a e f ,d} F_{b c g h i ,j} F_{e f g h i ,j}\nn\\&&\qquad\qquad\qquad\quad+6 F_{a b c ,d} F_{a b c ,d} F_{e f g h i ,j} F_{e f g h i ,j}+360 F_{a b c ,d} F_{a e f ,g} F_{b c h i j ,g} F_{e f h i j ,d}\nn\\&&\qquad\qquad\qquad\quad-360 F_{a b c ,d} F_{a e f ,g} F_{b c h i j ,d} F_{e f h i j ,g}+360 F_{a b c ,d} F_{a b e ,f} F_{c g h i j ,d} F_{e g h i j ,f}\nn\\&&\qquad\qquad\qquad\quad-36 F_{a b c ,d} F_{a b c ,e} F_{f g h i j ,d} F_{f g h i j ,e}+\epsilon_{a b c d e f g h i j} \bigg(3 F_{a b m,p} F_{k l m,n} F_{c d e k l,p} F_{f g h i j,n}\nn\\&&\qquad\qquad\qquad\quad-3 F_{a b m,n} F_{k l m,p} F_{c d e k l,p} F_{f g h i j,n}-3 F_{a b m,p} F_{k l m,p} F_{c d e k l,n} F_{f g h i j,n}\nn\\&&\qquad\qquad\qquad\quad+3 F_{a l m,n} F_{k l m,p} F_{b c d e k,p} F_{f g h i j,n}+2 F_{a b c,p} F_{k l m,p} F_{d e k l m,n} F_{f g h i j,n}\bigg)
\bigg]\labell{epsF5F5F3F3}
\eeqa
 The above couplings satisfy the self-duality condition $F^{(5)}=\star F^{(5)}$. Since the number of indices are too many, we could not find  all contraction of structure $(\prt F^{(5)})^2(\prt F^{(3)})^2$. So we could not performed the algorithm given above equation \reef{F3F3F3F3} to reduce the number of couplings to the minimum number. 

Note that under dimensional reduction on a circle, the P-odd couplings in \reef{epsF5F5F3F3} produce no term in which each RR field strength carries one Killing index. As a result, the   P-odd couplings in above equation produce no couplings with structure $\epsilon_{10} (F^{(4)})^2(F^{(2)})^2$ under T-duality. This is consistent with the fact that there is no such  coupling in \reef{F4F4F2F2}.

\subsection{$\prt F^{(n)}\prt F^{(n)}\prt F^{(n-4)}\prt F^{(n-4)}$ couplings}

Since the minimum rank of the RR field strength is 1, there is only one type of couplings in this section, \ie $n=5$. The effective action \reef{IIB} produces the following $s$-channel amplitude:
\beqa 
A_s&=&\left[\tV_{F_1^{(5)}F_2^{(5)}h}\right]^{\mu \nu}\left[\tG_{b} \right]_{\mu \nu, \lambda\rho}\left[\tV_{hF_3^{(1)}F_4^{(1)}}\right]^{\lambda\rho} \nn
\eeqa
In this case, one   can easily observe that there is no amplitude in   $u$- and  $ t $-channel. Therefore, the total amplitude comes from the $s$-channel which produces the following couplings in the string frame:
\beqa
&&\frac{\gamma}{2^8.5.3^2\kappa^2} \int d^{10}x\, e^{2\phi_0}\sqrt{-G} \, \bigg[240 F_{a,b} F_{c,d} F_{a e f g h ,c} F_{b e f g h ,d}\bigg]\nn
\eeqa
where we have also used the algorithm given above equation \reef{F3F3F3F3} to  reduce the number of couplings to minimum number. By imposing the self-duality transformation \reef{self-duality} on the above coupling, we obtain the following couplings for the self-dual RR form:
\beqa
S &\supset&\frac{\gamma}{2^9.5.3^2\kappa^2}  \int d^{10}x\, e^{2\phi_0}\sqrt{-G} \, \bigg[240 F_{a,b} F_{c,d} F_{a e f g h ,c} F_{b e f g h ,d}\labell{5511}\\&&-\epsilon_{a b c d e f g h i m} \bigg(F_{j,l} F_{k,m} F_{a b c d e,l} F_{f g h i j,k}
+F_{j,l} F_{k,m} F_{a b c d e,k} F_{f g h i j,l}-F_{k,j} F_{k,m} F_{a b c d e,l} F_{f g h i j,l}\bigg)\bigg]\nn
\eeqa
 where we have also  reduced the P-even couplings to minimum number.

\subsection{$\prt F^{(n)}\prt F^{(n-2)}\prt F^{(n-2)}\prt F^{(n-4)}$ couplings}

In this case also there is only one type of couplings, \ie $n=5$.  There is  no   Feynman amplitude in the $s$-channel. The amplitude in the $u$-channel is given as
\beqa
A_u&=&\left[\tV_{F_1^{(5)}F_3^{(3)}b}\right]^{\mu \nu}\left[\tG_{b} \right]_{\mu \nu, \lambda\rho}\left[\tV_{bF_2^{(3)}F_4^{(1)}}\right]^{\lambda\rho}+\left[\tV_{\epsilon_{10}F_1^{(5)}F_3^{(3)}b}\right]^{\mu \nu}\left[\tG_{b} \right]_{\mu \nu, \lambda\rho}\left[\tV_{bF_2^{(3)}F_4^{(1)}}\right]^{\lambda\rho} \nn\eeqa
The   $t$-channel amplitude is the same as $A_u$ in which the particle labels of the external RR fields are interchanged. 
Summing these   two contributions, one finds the amplitude produces the following P-even and P-odd couplings:
\beqa
 S&\supset&-\frac{\gamma}{2^6.5.3^2\kappa^2} \int d^{10}x\,e^{2\phi_0}\sqrt{-G} \, \bigg[120 F_{c d e f g ,h} F_{a ,b} F_{a c d ,b} F_{e f g ,h} -120 F_{c d f g h ,b} F_{a ,b} F_{a c d ,e} F_{f g h ,e}  \nn\\&&\qquad\qquad\qquad\qquad-120 F_{c d f g h ,e} F_{a ,b} F_{a c d ,e} F_{f g h ,b} +\epsilon_{a b c d e g h i j k} \bigg(F_{a b c d e,l} F_{f,m} F_{f g h,l} F_{i j k,m}\nn\\&&\qquad\qquad\qquad\qquad+F_{a b c d e,l} F_{f,l} F_{f g h,m} F_{i j k,m}-F_{a b c d e,l} F_{f,m} F_{f g h,m} F_{i j k,l} \bigg)\bigg]\labell{F5F1F3F3}
\eeqa
  Applying the self-duality condition \reef{self-duality} on above couplings and using the identity \reef{iden}, we have found  they are invariant.  
We have also    reduced the P-even couplings above to the minimum number. 

One may consider couplings with structure $\prt F^{(n)}\prt F^{(n-4)}\prt F^{(n-4)}\prt F^{(n-2)}$. In this case there is one possibility, \ie $n=5$. However, the type IIB supergravity indicates that the vertices in the Feynman amplitudes  are zero. So there is no such coupling at order $\alpha'^3$.

\subsection{$\prt F^{(n)}\prt F^{(n)}\prt F^{(n)}\prt F^{(n-2)}$ couplings}

There are three possibilities in this case, \ie $n=3,4,5$. However, the type IIB supergravity \reef{IIB} indicates that the Feynman amplitudes in $s$-, $t$- and $u$-channels are zero for $n=3,5$.  For the case $n=4$, the type IIA supergravity \reef{IIA} indicates that the amplitude in the $s$-channel  is given as 
\beqa 
A_s&=&\left[\tV_{\epsilon_{10}F_1^{(4)}F_2^{(4)}b}\right]^{\mu \nu}\left[\tG_{b} \right]_{\mu \nu, \lambda\rho}\left[\tV_{bF_3^{(4)}F_4^{(2)}}\right]^{\lambda\rho}\nn
\eeqa
The amplitude in the  $u$-channel is the same as $A_s$ in which the particle labels of the external RR fields are interchanged, \ie $A_u= A_s(2 \leftrightarrow 3)$. Similarly, the amplitude in the  $t$-channel is the same as $A_u$ in which the particle labels of the external RR fields are interchanged, \ie $A_t= A_u(1 \leftrightarrow 2)$. Replacing the appropriate vertices and propagators in the amplitudes, one finds the kinematic factors which produces the following  couplings in the string frame:
\beqa
S &\supset& \frac{\gamma}{2^9.3^2\kappa^2} \int d^{10}x\, e^{2\phi_0}\sqrt{-G} \, \epsilon_{a b c d e f g h i j}\bigg[ 2 F_{i j m n ,k} F_{e f g h ,l} F_{a b c d ,k} F_{m n ,l}\nn\\
&&\qquad\qquad\qquad\qquad\qquad\qquad\qquad\quad -F_{i j m n ,k} F_{e f g h ,l} F_{a b c d ,l} F_{m n ,k} \bigg]\labell{epsF4F4F4F2}
\eeqa
 which has only P-odd couplings.
  Note that under dimensional reduction on a circle, the above  couplings   produce no term  in which the  RR four-forms each  carries one Killing index and the RR two-form carries no Killing index. As a result, they produce   no couplings with structure $\epsilon_{10} (F^{(3)})^4 $ under T-duality. This is consistent with the fact that there is no such  couplings in \reef{F3F3F3F3}.

One may consider couplings with structure $\prt F^{(n)}\prt F^{(n)}\prt F^{(n)}\prt F^{(n-4)}$. In this case there is one possibility, \ie $n=5$. However, the type IIB supergravity indicates that the vertices in the Feynman amplitudes  are zero. So there is no such coupling at order $\alpha'^3$. It is also consistent with the T-duality of the couplings in \reef{epsF4F4F4F2}. In fact   the RR two-form in \reef{epsF4F4F4F2} does not contract with the Levi-Civita tensor, so under the dimensional reduction there is no coupling in which the RR four-forms carries no Killing index and the RR two-form carries one Killing index. As a result, the couplings \reef{epsF4F4F4F2} produce no terms with structure $\prt F^{(5)}\prt F^{(5)}\prt F^{(5)}\prt F^{(1)}$.

\section{Consistency with dualities}\label{ Duality approach} 

We have found all different couplings of four RR states at order $\alpha'^3$ in the previous section. In this section, we would like to show that these  couplings  are related to the standard four NSNS couplings  under    S-duality and T-duality transformations.  
We begin with the couplings in section 2.1. Our starting point in this case is the coupling $ F^{(3)}F^{(3)}F^{(3)}F^{(3)} $ which can be found by making the $H^4$ couplings to be S-duality invariant. The $ H^4 $ couplings on the other hand can be derived from the coupling $ t_8 t_8 R^4 $ in \reef{t8t8} by extending the Riemann curvature to the generalized Riemann curvature  \cite{Gross:1986mw}, \ie 
\beqa
{R}_{ab}{}^{cd}\rightarrow \bar{R}_{ab}{}^{cd}= R_{ab}{}^{cd}-\frac{\kappa}{\sqrt{2}}\eta_{[a}{}^{[c}\phi_{;b]}{}^{d]}+  2e^{-\phi_0/2}H_{ab}{}^{[c;d]}\, ,\labell{trans}
\eeqa
where the bracket notation is defined as $H_{ab}{}^{[c;d]}=\frac{1}{2}(H_{ab}{}^{c;d}-H_{ab}{}^{d;c})$ and $\phi_0$ is the constant dilaton background and the semicolon symbol denotes the covariant derivative. The resulting action has the following form in the Einstein frame  \cite{Gross:1986mw}:
\beqa
S &\!\!\!\!\!\supset\!\!\!\!\!&  \frac{\gamma}{\kappa^2}\int d^{10}x \sqrt{-G} e^{-3\phi_0/2} \bigg[\bar{R}_{hkmn}\bar{R}_{krnp}\bar{R}_{rsqm}\bar{R}_{shpq}+\frac{1}{2}\bar{R}_{hkmn}\bar{R}_{krnp}\bar{R}_{rspq}\bar{R}_{shqm}\labell{Y3}\\
&&\qquad\qquad\qquad\qquad\qquad-\frac{1}{2}\bar{R}_{hkmn}\bar{R}_{krmn}\bar{R}_{rspq}\bar{R}_{shpq} -\frac{1}{4}\bar{R}_{hkmn}\bar{R}_{krpq}\bar{R}_{rsmn}\bar{R}_{shpq}\nonumber\\
&&\qquad\qquad\qquad\qquad\qquad+\frac{1}{16}\bar{R}_{hkmn}\bar{R}_{khpq}\bar{R}_{rsmn}\bar{R}_{srpq}+\frac{1}{32}\bar{R}_{hkmn}\bar{R}_{khmn}\bar{R}_{rspq}\bar{R}_{srpq}\bigg]\nn
\eeqa
  The     couplings of four $H$ can be read from the above action. 
Invariance of this action under   S-duality transformations in type IIB theory requires the couplings of four $F^{(3)}$ to be the same as the couplings of four $H$, \ie the couplings in the string frame are 
\beqa 
S &\!\!\!\!\!\supset\!\!\!\!\!&  \frac{16 \gamma}{\kappa^2}\int d^{10}x \sqrt{-G} e^{2\phi_0} \bigg[F_{hk[m;n]}F_{kr[n;p]}F_{rs[q;m]}F_{sh[p;q]}+\frac{1}{2}F_{hk[m;n]}F_{kr[n;p]}F_{rs[p;q]}F_{sh[q;m]}\nonumber\\
&&\qquad\qquad\quad-\frac{1}{2}F_{hk[m;n]}F_{kr[m;n]}F_{rs[p;q]}F_{sh[p;q]} -\frac{1}{4}F_{hk[m;n]}F_{kr[p;q]}F_{rs[m;n]}F_{sh[p;q]}\nonumber\\
&&\qquad\qquad\quad+\frac{1}{16}F_{hk[m;n]}F_{kh[p;q]}F_{rs[m;n]}F_{sr[p;q]}+\frac{1}{32}F_{hk[m;n]}F_{kh[m;n]}F_{rs[p;q]}F_{sr[p;q]}\bigg].\labell{Y333}
\eeqa 
Writing the above couplings and the couplings $(F^{(3)})^4$  in \reef{F3F3F3F3} in terms of independent variables, we have found that they are exactly identical.

We use the following steps   on the couplings \reef{Y333}   to find the couplings with structure $(F^{(n)})^4 $ for $ n=2,1 $:  We first use the dimensional reduction on the couplings \reef{Y333}  and keep the terms with structure $(F^{(2)}_{y})^4 $ where the index $y$ is the Killing index. Under the linear T-duality transformations, the RR field strength $ F^{(n)}_{y} $ transforms to $ F^{(n-1)} $ with no Killing index. Therefore, under the T-duality transformation the above couplings transform to the couplings with structure $(F^{(2)})^4 $ in type IIA theory. Performing the same steps once more, we have found the couplings with structure $(F^{(1)})^4 $ in type IIB theory. We have checked that these couplings are exactly equal to the corresponding couplings in \reef{F3F3F3F3} when we write them in terms of independent variables.

Now consider the couplings in the dimensional reduction of  \reef{Y333} in which the RR three-forms carry no Killing index. Under the T-dality, they transforms to the  couplings  with structure $(F_y^{(4)})^4$ in type IIA theory. We compare these couplings with the couplings with structure $(F_y^{(4)})^4$ in the dimensional reduction of the couplings \reef{F4F4F4F4}. Writing both set of couplings in terms of independent variables, we have found exact agreement.

To show that the coupling with structure $(F^{(5)})^4 $ in \reef{epsF5F5F5F5} are consistent with  dualities, we note that RR five-form field strength is invariant under the S-duality. We already pointed out that the P-odd couplings in \reef{epsF5F5F5F5} are consistent with T-duality. To verify that the P-even couplings in  \reef{epsF5F5F5F5} are consistent with T-duality, we consider the couplings in the dimensional reduction of  \reef{epsF5F5F5F5} in which the RR five-forms each carries one Killing index, \ie $(F_y^{(5)})^4$. Under the T-duality, they transforms to the couplings with structure  $(F^{(4)})^4$. We compare them with the couplings in \reef{F4F4F4F4}. Writing both set of couplings in terms of independent variables, we have found exact agreement.

We now compare the couplings in section 2.2 with dualities. Our starting point in this case is the couplings with structure   $F^{(5)}F^{(5)}F^{(3)}F^{(3)}$. Using the consistency of the couplings \reef{Y3} with S-duality and  T-duality, the couplings with structure  $F^{(5)}F^{(5)}HH$ have been found in  \cite{Garousi:2013lja}.  Under the S-duality, the RR five-form is invariant and B-field strength $H$ transforms to the RR three-form field strength. So the  consistency of the couplings found in \cite{Garousi:2013lja} with  S-duality, requires  the following couplings in the string frame:  
\beqa
S &\supset&\frac{\gamma}{\kappa^2} \int d^{10}x\, e^{2\phi_0}\sqrt{-G}\bigg[-\frac{2}{3} F_{h r s t u,n} F_{q r s t u,m} F_{k n p,h} F_{m p q,k}\nn\\&&+\frac{2}{3} F_{n q s t u,h} F_{p q s t u,m} F_{k n r,h} F_{m p r,k}-\frac{1}{90} F_{n q s t u,h} F_{n q s t u,k} F_{m p r,h} F_{m p r,k}\nonumber\\&&+\frac{1}{6} F_{h q s t u,n} F_{k q s t u,m} F_{m p r,k} F_{n p r,h}+\frac{1}{6} F_{h q s t u,m} F_{k q s t u,n} F_{m p r,k} F_{n p r,h}\nonumber\\&&+\frac{4}{9} F_{h k m n u,t} F_{p q r t u,n} F_{h k m,s} F_{p q r,s}+\frac{1}{3} F_{n q r t u,h} F_{n p s t u,k} F_{h q r,m} F_{k p s,m}\nonumber\\&&-\frac{1}{3} F_{n q r t u,h} F_{m r s t u,k} F_{n p q,k} F_{m p s,h}-\frac{4}{3} F_{m n p t u,k} F_{n r s t u,h} F_{k p q,m} F_{h q s,r}\nonumber\\&&+4 F_{n q r t u,h} F_{m p s t u,k} F_{h k r,n} F_{p q s,m}+\frac{1}{3} F_{m n p t u,h} F_{m q s t u,h} F_{n p r,k} F_{q r s,k}\bigg].\labell{F5F3}
\eeqa
To compare them with the couplings in \reef{epsF5F5F3F3}, we have to impose the self-duality transformation \reef{self-duality} on the above couplings. Using the identity \reef{iden} to write the multiple of two Levi-Civita tensors in terms of metric, we find  two types of terms. One type  has terms with no Levi-Civita tensor which is the same as \reef{F5F3} up to the overall factor of $\frac{1}{2}$. The other type  has terms with one Levi-Civita tensor. This part,  in the momentum space,  produces terms with zero or one Mandelstam variables which are not consistent with the superstring theory amplitudes \cite{Garousi:2012yr,Barreiro:2012aw}.  This indicates that the couplings \reef{F5F3} must have some P-odd couplings which are not related  to the couplings \reef{Y3} by   the string dualities.  In fact, the corresponding Feynman amplitude in section 2.2 has P-odd couplings   even before imposing the self-duality transformation. 

The P-even couplings \reef{F5F3} and their P-odd partners  must be consistent with T-duality before or after imposing the self-duality transformation  because they are not produced by type IIB supergravity which is off for the RR five-form field strength. In particular, under the dimensional reduction of \reef{F5F3}, the couplings with structure $(F_y^{(5)})^2(F_y^{(3)})^2$ transforms under T-duality to the couplings with structure $(F^{(4)})^2(F^{(2)})^2$ in \reef{F4F4F2F2}. This indicates that the self-duality transformation of the action \reef{F5F3} and its P-odd  partner  should  produce the same couplings as \reef{F5F3}. The transformation of \reef{F5F3} under the self-duality \reef{self-duality} produces the same couplings with the overall factor  $\frac{1}{2}$ and some P-odd couplings. The other factor of $\frac{1}{2}$ must then be reproduced by self-duality transformation of the P-odd terms.

One may try to find the P-odd partner  of \reef{F5F3} by considering all contractions with structure $\epsilon^{(10)}(F^{(5)})^2(F^{(3)})^2$ with unknown coefficients and fix them by requiring them to produce the above  factor of $\frac{1}{2}$ under the self-duality transformation and requiring them to   produce no term with zero or one Mandelstam variables in the momentum space \cite{Garousi:2012yr,Barreiro:2012aw}.  However, there are too many    such contractions, so we do not try to find the P-odd partner of the couplings  \reef{F5F3} in this paper. We have written the  couplings \reef{F5F3}   and the  P-even part of couplings \reef{epsF5F5F3F3}   in terms of independent variables and found  exact agreement.  

Using the dimensional reduction on the   couplings \reef{F5F3} in type IIB theory, one can find the couplings with structure $(F^{(5)}_{y})^2(F^{(3)}_{y} )^2$. Under T-duality they  transform to the couplings with structure $(F^{(4)})^2(F^{(2)})^2$ in type IIA theory. Repeating these steps   on the couplings  $(F^{(4)})^2(F^{(2)})^2$, one can find   the couplings with structure $(F^{(3)})^2(F^{(1)})^2$ in type IIB theory. Writing these couplings and the corresponding couplings in \reef{F4F4F2F2} in terms of independent variables, we have again found exact agreement between the two set of couplings.

We now compare the couplings in section 2.3 with dualities. There is only one coupling in this section, \ie the couplings with structure   $(F^{(5)})^2(F^{(1)})^2$. Such couplings have been found in  \cite{Bakhtiarizadeh:2013zia} by imposing the S-duality on the couplings of two RR five-form and two dilatons (see  eq. (69) in \cite{Bakhtiarizadeh:2013zia}). Alternatively, the couplings with structure  $(F^{(5)})^2(F^{(1)})^2$ can be found by imposing the dualities on the couplings \reef{F5F3}. To this end, we use the dimensional reduction on the couplings with structure   $(F^{(4)})^2(F^{(2)})^2$  that we have found in the above paragraph, and consider terms with structure  $(F^{(4)})^2 (F^{(2)}_y)^2$. Then under T-duality they transform to the couplings with structure $(F^{(5)}_{y})^2(F^{(1)})^2$. Converting the Killing index to a complete space-time index and taking the symmety factors into account, one finds the couplings with structure $(F^{(5)})^2(F^{(1)})^2$ without any ambiguity because it is impossible to have couplings in which the RR five-forms do not contract with each other. These couplings are the same as the couplings found in \cite{Bakhtiarizadeh:2013zia}, \ie in the string frame they are
\beqa
S &\supset & \frac{2}{3}\frac{\gamma}{\kappa^2} \int d^{10}x\sqrt{-G}\, e^{2\phi_0} \bigg[  F_{a e f g h ,c} F_{b e f g h ,d} F_{a ,b} F_{c ,d} \bigg]
\eeqa
 Transforming  the above couplings under the self-duality \reef{self-duality}, we have found  couplings which are identical to the couplings in \reef{5511} after writing both sets in terms of independent variables. This indicates that there is no P-odd coupling in the above action.  As in the field theory section 2.3, the P-odd part in the self-dual action comes only from the self-duality transformation.

To compare the couplings in section 2.4 with dualities, we consider  the couplings with structure  $ F^{(5)}F^{(1)}HH $ which have been found in \cite{Bakhtiarizadeh:2013zia} by imposing dualities on the couplings in \reef{Y333}. The  S-duality invariant of these couplings produces among other things the couplings with structure $F^{(5)}F^{(1)}F^{(3)}F^{(3)}$, \ie
\beqa
S &\supset&- \frac{\gamma}{\kappa^2}  \int d^{10}x\sqrt{-G}e^{2\phi_0}\big[ 8 F_{h,k} F_{{mnpqr},s} F_{{hpq},m} F_{{krs},n}\labell{F1F5}\\&&\qquad\qquad\qquad\quad+4 F_{h,k} F_{{kmnpq},r} F_{{mns},h} F_{{pqr},s}-2 F_{h,k} F_{{kmnpq},h} F_{{mns},r} F_{{pqr},s}
\big]\nn
\eeqa
Note that these couplings are only P-even. Transforming  them under the self-duality \reef{self-duality}, one finds P-even and P-odd couplings. Writing them  in terms of independent variables, we have found they are exactly equal to the couplings in \reef{F5F1F3F3}. It is important to note that  the field theory couplings in section 2.4 have P-odd  part even before imposing the self-duality transformation, whereas the above couplings   have no P-odd part. This indicates the field theory couplings \reef{F5F1F3F3} are   consistent with the duality transformations of \reef{Y3}, \ie \reef{F1F5},  only after imposing the self-duality transformation on \reef{F1F5}.

Finally, the  P-odd couplings in section 2.4 should be related to the P-odd couplings in \reef{epsF4F4F4F2} under T-duality. Under the dimensional reduction, the latter couplings produce among other things, the couplings with structure $\epsilon^{(10)}_yF^{(4)}(F^{(4)}_y)^2F_y^{(2)}$. Under T-duality, they transform to the couplings with structure $\epsilon^{(10)}_yF^{(5)}_y(F^{(3)})^2F^{(1)}$. Completing the Killing index to the full spacetime index, one finds the couplings with structure $ \epsilon_{10} F^{(5)} (F^{(3)})^2F^{(1)}$. Writing the resulting couplings and the P-odd couplings in \reef{F5F1F3F3} in terms of independent variables, we have found exact agreement. 

\section{Consistency with string  amplitudes}\label{String theory amplitude} 

 In this section, we are going to calculate the kinematic factor $\cK$ directly in the type II superstring theory and compare it with the couplings found in section 2.

The  tree-level scattering amplitude of four RR states in the RNS formalism \cite{Friedan:1985ge}     is given by the correlation function of their corresponding vertex operators on the sphere world-sheet. Since the background superghost charge of the sphere is $Q_{\phi}=2$, one has to choose the vertex operators in the appropriate pictures to produce the compensating charge $Q_{\phi}=-2$.  We choose the RR vertex operators in $(-1/2,-1/2)$  picture.  The amplitude   is given by the following correlation function \cite{Friedan:1985ge}:
\beqa
{\cal A}\sim\int \prod_{i=1}^{4} d^2 z_{i} \; \left\lan \prod_{j=1}^{4} V_{RR}^{(-1/2,-1/2)}(z_{j},\bar{z}_{j})\right\ran, \labell{amp1}
\eeqa
where the vertex operators are
\beqa
V_{RR}^{(-1/2,-1/2)}(z_{j},\bar{z}_{j})&=&(P_{\mp}\Gamma_{j(n)})^{AB}:e^{-\phi(z_{j})/2}S_{A}(z_{j})e^{ik_{j}\cdot X(z_{j})}:\nn\\&& \qquad\qquad\quad\times e^{-\tphi(\bar{z}_{j})/2}\tS_{B}(\bar{z}_{j})e^{ik_{j}\cdot \tX({\bar{z}_{j}})}:,\labell{vert}
\eeqa
where $j=1,\cdots4$ and the indices $A,B,\cdots$ are the Dirac spinor indices and  $P_\mp=\frac{1}{2}(1\mp\gamma_{11})$ is the chiral projection operators which make the calculation of the gamma matrices to be with the full $32\times 32$ Dirac matrices of the ten dimensions.  The RR field strength appears in the definition of $\Gamma_{i(n)} $  as 
\beq
\Gamma_{i(n)}=\frac{a_n}{n!} F_{i \mu_1\cdots\mu_n}\,\gamma^{\mu_1\cdots\mu_n},\labell{self}
\eeq
where the factor $a_n=-1$ in the type IIA theory and $a_n=i$ in the type IIB theory \cite{Garousi:1996ad}. There is ambiguity in choosing the  chiral projection operator in the vertex operator \reef{vert}, \eg $P_-$ or $P_+$.  As we will see, this makes it difficult to  confirm the P-odd couplings in section 2 with the string theory scattering amplitudes.  The normalization   of the amplitude \reef{amp1} in which we are not interested in this section,     may be fixed after fixing the conformal symmetry of the integrand.

Substituting the vertex operators \reef{vert} into \reef{amp1}, and using the fact that there is no correlation between holomorphic and anti-holomorphic for the sphere world-sheet, one can separate the amplitude to the holomorphic and the anti-holomorphic parts as   
\beqa
{\cal A} \sim (P_{-} \Gamma_{1(n)})^{AB}(P_{-} \Gamma_{2(m)})^{CD}(P_{-} \Gamma_{3(p)})^{EF}(P_{-} \Gamma_{4(q)})^{GH}\int \prod_{i=1}^{4} d^2 z_{i}\; I_{ACEG}\otimes\tilde{I}_{BDFH},\labell{amp2}
\eeqa
where the holomorphic part is   
\beqa
I_{AC}^{\mu \alpha}&=&\left\lan:e^{-\phi(z_{1})/2}:e^{-\phi(z_{2})/2}:e^{-\phi(z_{3})/2}:e^{-\phi(z_{4})/2}:\right\ran\nn\\&&\times\left\lan:e^{ik_{1}\cdot X(z_{1})}:e^{ik_{2}\cdot X(z_{2})}:e^{ik_{3}\cdot X(z_{3})}:e^{ik_{4}\cdot X(z_{4})}:\right\ran\nn\\&&\times\left\lan:S_{A}(z_{1}):S_{C}(z_{2}):S_{E}(z_{3}):S_{G}(z_{4}):\right\ran,\labell{right}
\eeqa
and the anti-holomorphic part $\tilde{I}_{BDFH}$ is given by similar expression.

Using the standard propagators and the correlation function of four spin operators \cite{Friedan:1985ge}, one can perform the correlatos in \reef{amp2}. Using the on-shell relations and  the conservation of momentum,  one can  check that the integrand of the   amplitude is invariant under $SL(2,{R})\times SL(2,{R})$ transformations which is the conformal symmetry of the   $z$-plane. Fixing this symmetry by setting $z_{1}=0,z_{2}\equiv z,z_{3}=1$ and $z_{4}=\infty$ and normalizing the amplitude, one can write  it in the string frame form of \reef{contact} in which the kinematic factor is  
\beqa
\cK&=&(P_{\mp} \Gamma_{1(n)})^{AB}(P_{\mp} \Gamma_{2(m)})^{CD}(P_{\mp} \Gamma_{3(p)})^{EF}(P_{\mp} \Gamma_{4(q)})^{GH} K_{ACEG}\otimes\tilde K_{BDFH}.\labell{kin0}
\eeqa
where the kinematic factor in the holomorphic part is
\beqa
K_{ACEG}&=& -\frac{1}{8}\bigg[t (\gamma^{\mu}C^{-1})_{AC}(\gamma_{\mu}C^{-1})_{EG}-s (\gamma^{\mu}C^{-1})_{AG}  (\gamma_{\mu}C^{-1})_{CE} \bigg].\labell{kin1}
\eeqa
and the kinematic factor in the anti-holomorphic part is  similar to the above expression.  

One may use the  KLT prescription \cite{Kawai:1985xq} to calculate the sphere-level scattering amplitude of closed string states from the corresponding disk-level scattering amplitude of open string states. According to the KLT prescription, the sphere-level amplitude of four closed string states is given by
\beqa
{\cal A}&\sim&  \sin(\alpha'\pi k_2\inn k_3/2)A_{\rm open}(s/8,t/8)\otimes\tA_{\rm open}(t/8,u/8),\labell{KLT}
\eeqa
where $A_{\rm open}(s/8,t/8)$ is the disk-level scattering amplitude of four open string states in the $s-t$ channel which has been calculated in \cite{Schwarz:1982jn},
\beqa
A_{\rm open}(s/8,t/8)&\sim& \frac{\Gamma(-s/8)\Gamma(-t/8)}{\Gamma(1+u/8)}K, \labell{open}
\eeqa
where the Mandelstam variables are the same as in the closed string amplitude. The open string kinematic factor $K$  depends on the momentum and the polarization of the external states \cite{Schwarz:1982jn}.  

To find the sphere-level scattering amplitude of four RR states, one has to consider the disk-level scattering amplitude of four R states. The kinematic factor for this case is \cite{Schwarz:1982jn}
\beqa
K(u_1,u_2,u_3,u_4)&=& -\frac{1}{8}\bigg[t \bar{u}_1^{A} (\gamma^{\mu}C^{-1})_{AC}u_2^{C}\bar{u}_3^{E}(\gamma_{\mu}C^{-1})_{EG}u_4^{G}\nn\\
&&-s \bar{u}_1^{A} (\gamma^{\mu}C^{-1})_{AG} u_4^{G}\bar{u}_2^{C}  (\gamma_{\mu}C^{-1})_{CE} u_3^{E}\bigg]\labell{openkin}
\eeqa
where $ u_{i} $ with $ i=1,\cdots,4 $ are the spinor polarizations. They satisfy the following on-shell relations  
\beqa
k_{i}^{2}=0,\qquad (\gamma\inn k_{i}C^{-1})_{AB}u_i^{B}=0.\labell{on-shell}
\eeqa
Using these relations, one can write the open string kinematic factor \reef{openkin} in terms of the holomorphic kinematic factor \reef{kin1} as 
\beqa
K(u_1,u_2,u_3,u_4)&=& -4i\sqrt{2}u_1^{A}u_2^{C}u_3^{E}u_4^{G} K_{ACEG}.\labell{trans2}
\eeqa
Similarly for the antiholomorphic part, \ie
\beqa
\tilde K(\tilde u_1,\tilde u_2,\tilde u_3,\tilde u_4)&=&-4i\sqrt{2}\tilde u_1^{B}\tilde u_2^{D}\tilde u_3^{F}\tilde u_4^{H} \tilde K_{BDFH}.\nn
\eeqa
Using the above relations and $\Gamma(x)\Gamma(1-x)=\pi/\sin(\pi x)$,  and  substituting the following relations in \reef{KLT}     
\beqa
u_1^{A}\otimes \tilde u_1^{B}&\rightarrow& (P_{\mp} \Gamma_{1(n)})^{AB},\nn\\
u_2^{C}\otimes \tilde u_2^{D}&\rightarrow& (P_{\mp} \Gamma_{2(m)})^{CD},\nn\\
u_3^{E}\otimes \tilde u_3^{F}&\rightarrow& (P_{\mp} \Gamma_{1(n)})^{EF},\nn\\
u_4^{G}\otimes \tilde u_4^{H}&\rightarrow& (P_{\mp} \Gamma_{2(m)})^{GH},\labell{trans1}
\eeqa
one can write it as the string frame form of \reef{contact} with the kinematic factor  \reef{kin0}, as expected. While the open string kinematic factor \reef{openkin} is the final result for the S-matrix element of four open string spinors, the closed string kinematic factor \reef{kin0} is not yet the final result. The Dirac matrices in the kinematic factor  appear in   trace operators which should then be evaluated explicitly to find the final kinematic factor of the closed string amplitude. 

The closed string kinematic factor \reef{kin0} has four different terms, each one has one of the following factors:
\beqa
T_1&=&(P_{\mp} \Gamma_{1(n)})^{AB}(P_{\mp} \Gamma_{2(m)})^{CD}(P_{\mp} \Gamma_{3(p)})^{EF}(P_{\mp} \Gamma_{4(q)})^{GH}\nn\\&&\times(\gamma^{\mu}C^{-1})_{AC}(\gamma_{\mu}C^{-1})_{EG}(\gamma^{\nu}C^{-1})_{BD}(\gamma_{\nu}C^{-1})_{FH},\nn\\ \nn\\
T_2&=&(P_{\mp} \Gamma_{1(n)})^{AB}(P_{\mp} \Gamma_{2(m)})^{CD}(P_{\mp} \Gamma_{3(p)})^{EF}(P_{\mp} \Gamma_{4(q)})^{GH}\nn\\&&\times(\gamma^{\mu}C^{-1})_{AG}(\gamma_{\mu}C^{-1})_{CE}(\gamma^{\nu}C^{-1})_{BD}(\gamma_{\nu}C^{-1})_{FH},\nn\\ \nn\\
T_3&=&(P_{\mp} \Gamma_{1(n)})^{AB}(P_{\mp} \Gamma_{2(m)})^{CD}(P_{\mp} \Gamma_{3(p)})^{EF}(P_{\mp} \Gamma_{4(q)})^{GH}\nn\\&&\times(\gamma^{\mu}C^{-1})_{AC}(\gamma_{\mu}C^{-1})_{EG}(\gamma^{\nu}C^{-1})_{BH}(\gamma_{\nu}C^{-1})_{DF},\nn\\ \nn\\ 
T_4&=&(P_{\mp} \Gamma_{1(n)})^{AB}(P_{\mp} \Gamma_{2(m)})^{CD}(P_{\mp} \Gamma_{3(p)})^{EF}(P_{\mp} \Gamma_{4(q)})^{GH}\nn\\&&\times(\gamma^{\mu}C^{-1})_{AG}(\gamma_{\mu}C^{-1})_{CE}(\gamma^{\nu}C^{-1})_{BH}(\gamma_{\nu}C^{-1})_{DF}, \labell{tr0}
\eeqa
  Using the properties of the charge conjugation matrix and the Dirac matrices (see \eg appendix B. in \cite{Garousi:1996ad}), one can write the tensors  $T_1,\cdots, T_4$  in terms of the RR field strengths and the trace of the gamma matrices as
\beqa
T_1&=& \frac{(-1)^{\frac{1}{2}[m(m+1)+q(q+1)]}a_{n}a_{m}a_{p}a_{q}}{n!m!p!q!}F_{1\mu_1\cdots\mu_{n}}F_{2\nu_1\cdots\nu_{m}}F_{3\alpha_1\cdots\alpha_{p}}F_{4\beta_1\cdots\beta_{q}}\nn\\&&\Tr(P_\pm \gamma^{\mu} \gamma^{\mu_1\cdots\mu_{n}}\gamma^{\nu} \gamma^{\nu_1\cdots\nu_{m}})\Tr(P_\pm \gamma_{\mu} \gamma^{\alpha_1\cdots\alpha_{p}}\gamma_{\nu} \gamma^{\beta_1\cdots\beta_{q}}),\nn\\
T_2&=& \frac{(-1)^{\frac{1}{2}[m(m+1)+q(q+1)]}a_{n}a_{m}a_{p}a_{q}}{n!m!p!q!}F_{1\mu_1\cdots\mu_{n}}F_{2\nu_1\cdots\nu_{m}}F_{3\alpha_1\cdots\alpha_{p}}F_{4\beta_1\cdots\beta_{q}}\nn\\&&\Tr(P_\pm \gamma^{\nu} \gamma^{\nu_1\cdots\nu_{m}} \gamma^{\mu} \gamma^{\alpha_1\cdots\alpha_{p}}\gamma_{\nu} \gamma^{\beta_1\cdots\beta_{q}}\gamma_{\mu}\gamma^{\mu_1\cdots\mu_{n}}),\nn\\
T_3&=& \frac{(-1)^{\frac{1}{2}[n(n+1)+p(p+1)]}a_{n}a_{m}a_{p}a_{q}}{n!m!p!q!}F_{1\mu_1\cdots\mu_{n}}F_{2\nu_1\cdots\nu_{m}}F_{3\alpha_1\cdots\alpha_{p}}F_{4\beta_1\cdots\beta_{q}}\nn\\&&\Tr(P_\pm \gamma^{\nu} \gamma^{\nu_1\cdots\nu_{m}} \gamma^{\mu} \gamma^{\alpha_1\cdots\alpha_{p}}\gamma_{\nu} \gamma^{\beta_1\cdots\beta_{q}}\gamma_{\mu}\gamma^{\mu_1\cdots\mu_{n}}),\nn\\
T_4 &=& \frac{(-1)^{\frac{1}{2}[m(m+1)+q(q+1)]}a_{n}a_{m}a_{p}a_{q}}{n!m!p!q!}F_{1\mu_1\cdots\mu_{n}}F_{2\nu_1\cdots\nu_{m}}F_{3\alpha_1\cdots\alpha_{p}}F_{4\beta_1\cdots\beta_{q}}\nn\\&&\Tr(P_\pm \gamma^{\mu} \gamma^{\mu_1\cdots\mu_{n}}\gamma^{\nu}\gamma^{\beta_1\cdots\beta_{q}})\Tr(P_\pm \gamma_{\mu} \gamma^{\alpha_1\cdots\alpha_{p}}\gamma_{\nu} \gamma^{\nu_1\cdots\nu_{m}}).\labell{tr}
\eeqa
Using the above factors,     the closed string kinematic factor \reef{kin0} can be written as 
\beqa
{\cal K}=\frac{1}{64} \bigg[t^{2} T_1  -  st T_2  -  st T_3 + s^{2} T_4\bigg],\labell{kin2}
\eeqa
Performing the traces, one finds how the four  RR field strengths contract among themselves. Writing $t^2=16\alpha'^2 k_2\inn k_3 k_1\inn k_4$, $s^2=16\alpha'^2k_1\inn k_2 k_3\inn k_4$ and using the first relation in \reef{identities}, one   may write the kinematic factor \reef{kin2} in the form of  $(\prt F)^4$ in the spacetime which can then be compared with the couplings in section 2.

The scattering amplitude of four RR states in type II superstring theories have been also calculated in the Pure  spinor formalism in \cite{Policastro:2006vt}. The couplings of four RR field strengths at order $\alpha'^3$ have been found to be   
\beqa
S &\supset &\sum_{M,N,P,Q}v^{a_1\cdots a_{M};b_1\cdots b_{N};c_1\cdots c_{P};d_1\cdots d_{Q}}\partial_{i}\partial_{j}F_{a_1\cdots a_M} \partial^{i}\partial^{j} F_{b_1\cdots b_N} F_{c_1\cdots c_P} F_{d_1\cdots d_Q},
\eeqa
where the sum over $ M,\cdots ,Q $, run over even integers from zero to four for type IIA supergravity, and over odd integers from one to five for type IIB. The tensor $v$ is defined in terms of the trace of the gamma matrices as follows: 
\beqa
v^{a_1\cdots a_{M};b_1\cdots b_{N};c_1\cdots c_{P};d_1\cdots d_{Q}}&=&\nn \frac{32}{9}\frac{c_{M}c_{N}c_{P}c_{Q}}{M!N!P!Q!} \\&& \bigg[ \Tr(P_{\mp}\gamma^{a_1\cdots a_{M}}\gamma_q \gamma^{b_1\cdots b_{N}}\gamma_n \gamma^{c_1\cdots c_{P}}\gamma^q\gamma^{d_1\cdots d_{Q}}\gamma^n )\veps_{N}\veps_{Q}\nn\\
&&-\Tr(P_{\mp}\gamma^{a_1\cdots a_{M}}\gamma_q \gamma^{b_1\cdots b_{N}}\gamma_n)\Tr( P_{\mp}\gamma^{c_1\cdots c_{P}}\gamma^q\gamma^{d_1\cdots d_{Q}}\gamma^n )\veps_{N}\veps_{Q}\nn\\
&& -5 \, \Tr(P_{\mp}\gamma^{a_1\cdots a_{M}}\gamma_q \gamma^{c_1\cdots c_{P}}\gamma_n  \gamma^{b_1\cdots b_{N}}\gamma^q \gamma^{d_1\cdots d_{Q}}\gamma^n )\veps_{P}\veps_{Q}\nn\\
&& + 4\, \Tr(P_{\mp}\gamma^{a_1\cdots a_{M}}\gamma_q \gamma^{c_1\cdots c_{P}}\gamma_n)  \Tr(P_{\mp}\gamma^{b_1\cdots b_{N}}\gamma^q\gamma^{d_1\cdots d_{Q}}\gamma^n)\veps_{P}\veps_{Q}\nn\\
&&+\Tr(P_{\mp}\gamma^{a_1\cdots a_{M}}\gamma_q\gamma^{c_1\cdots c_{P}} \gamma_n \gamma^{d_1\cdots d_{Q}}\gamma^q\gamma^{b_1\cdots b_{N}}\gamma^n )\veps_{N}\veps_{Q}\bigg]\labell{v}
\eeqa
where $c_p^2=(-1)^{p+1}/16\sqrt{2}$ and $\veps_N=(-1)^{\frac{1}{2}N(N-1)}$. We have included the chiral projection operators in the traces, because the RR vertex operators that have been considered in  \cite{Policastro:2006vt} have no chiral projection operator. Using the RR vertex operator \reef{vert} instead, one has to consider $P_{\mp}$ inside the traces.

The $\gamma_{11}$ in the chiral projection operators has one Levi-Civita tensor. As a result, the kinematic factor \reef{kin2} and tensor $v$ in \reef{v} have terms with zero, one and two Levi-Civita tensors. Since there is ambiguity in the chiral projection operator in the vertex \reef{vert}, the signs of P-odd terms  in $T_1,\, T_2, \,T_3, \,T_4$  and in   tensor $v$ are ambitious. Therefor, we consider only the P-even terms in the kinematic factor \reef{kin2} and in  tensor $v$ . Moreover, we use  the identity \reef{iden} to write the two Levi-Civita tensors in them  in terms of metric. Using the  symbolic program for the manipulation the gamma matrices \cite{Gran:2001yh}, we have performed the traces in the kinematic factor \reef{kin2} and in the tensor $v$. Using on-shell relations, we have found that the P-even terms   in RNS   and in the Pure spinor formalisms at order $\alpha'^3$  are exactly identical. We have also found   these couplings are identical to various   P-even couplings in section 2.


{\bf Acknowledgments}: This work is supported by Ferdowsi University of Mashhad under grant 2/30889-1393/04/10.  

\appendix

\section{Three-point vertices and propagators}\label{propagators and vertices}

Using the supergravities  \reef{IIA} and \reef{IIB}, one  can read  the propagators of the NSNS fields and the three-point vertices for two on-shell RR states and one off-shell NSNS state that we need in evaluating the Feynman amplitudes in section 2.  The propagators are 
\begin{itemize}
\item Graviton propagator
\beqa
\left[\tG_h\right]_{\mu\nu,\lambda\rho}&=&
-\frac{i}{2k^2}\left(\eta_{\mu\lambda}\eta_{\nu\rho}+
\eta_{\mu\rho}\eta_{\nu\lambda}-\frac{1}{4}\eta_{\mu\nu}
\eta_{\lambda\rho}\right).\labell{Grapro}
\eeqa
\item B-field propagator
\beqa
\left[\tG_b\right]_{\mu\nu,\lambda\rho}&=&-\frac{ie^{\phi_0}}{2 k^2}\left(\eta_{\mu\lambda}\eta_{\nu\rho}
-\eta_{\mu\rho}
\eta_{\nu\lambda}\right).\labell{Bpro}
\eeqa
\item Dilaton propagator
\beqa
\tG_\phi &=&-\frac{i}{k^2}.\labell{Dilpro}
\eeqa
\end{itemize}
The vertices are the following:
\begin{itemize}
\item Two RR and one graviton\footnote{The  parentheses notation over indices means  symmetrization with a factor $\frac{1}{2}$.}
\beqa
\left[\tV_{F_1^{(n)}F_2^{(n)}h}\right]^{\lambda\rho}&=&  
\frac{ie^{\frac{(5-n)}{2}\phi_0}}{4\kappa \alpha n!}\left(2n\,F_{1}^{(\lambda}{}_{\nu_1\cdots\nu_{n-1}}F_2^{\rho)\nu_1\cdots\nu_{n-1}}  -
\eta^{\lambda\rho}\,F_ {1\nu_1\cdots\nu_{n}}F_2^{\nu_1\cdots\nu_{n}} \right) \labell{RRh}
\eeqa
\item Two RR and one B-field 
\beqa
\left[\tV_{F_1^{(n)}F_2^{(n-2)}b}\right]^{\lambda\rho}&=&- \frac{ie^{\frac{(5-n)}{2}\phi_0}}{4\kappa \alpha (n-2)!}F_{1}^{\lambda\rho}{}_{\nu_1\cdots\nu_{n-2}}F_2^{\nu_1\cdots\nu_{n-2}} \labell{RRB}
\eeqa 
\item Two RR and one dilaton 
\beqa
\tV_{F_1^{(n)}F_2^{(n)}\phi}&=&
- \frac{ie^{\frac{(5-n)}{2}\phi_0}}{4\sqrt{2}\kappa  n!}(5-n)\,F_ {1\nu_1\cdots\nu_{n}}F_2^{\nu_1\cdots\nu_{n}} \labell{RRdil}
\eeqa
 \item Two RR four-form, one B-field and  one Levi-Civita tensor
\beqa
\left[\tV_{\epsilon_{10}F_1^{(4)}F_2^{(4)}b}\right]^{\alpha  \beta}&=&-\frac{i }{1152\kappa}    \epsilon^{\alpha  \beta  \gamma  \delta  \epsilon  \zeta  \eta  \lambda  \mu  \nu} F_{1\gamma  \delta  \epsilon  \zeta } F_{2\eta  \lambda  \mu  \nu }.
\eeqa
 \item One RR five-form, one RR three-form, one B-field and  one Levi-Civita tensor
\beqa
\left[\tV_{\epsilon_{10}F_1^{(5)}F_2^{(3)}b}\right]^{\alpha  \beta}&=&-\frac{i}{2880\kappa}   \epsilon^{\alpha  \beta  \gamma  \delta  \epsilon  \zeta  \eta  \lambda  \mu  \nu } F_{1\zeta  \eta  \lambda  \mu  \nu} F_{2\gamma  \delta  \epsilon }.
\eeqa
 \end{itemize}

\end{document}